\documentclass[11pr]{article}

\usepackage[figuresright]{rotating}
\usepackage[dvips]{color}
\usepackage{amsmath,amssymb}
\usepackage[round]{natbib}
\usepackage{bm}

\def\bSig\mathbf{\Sigma}

\newcommand{\bmath}[1]{\mbox{\boldmath $\!#1\!$ \unboldmath}}

\begin{document}

\title{\bf Joint Clustering and Registration\\ of Functional Data}

\author{Yafeng Zhang and
Donatello Telesca \\
Department of Biostatistics,  University of California Los Angeles,\\ Los Angeles,
California, U.S.A.}

\maketitle

\centerline{Abstract}
\noindent Curve registration and clustering are fundamental tools in the analysis of functional data. While several methods have been developed and explored for either task individually, limited work has been done to infer functional clusters and register curves simultaneously.  We propose a hierarchical model for joint curve clustering and  registration. Our proposal combines a Dirichlet process mixture model for clustering of common shapes, with a reproducing kernel representation of phase variability for registration.   We show how inference can be carried out applying standard posterior simulation algorithms and compare our method to several alternatives in both engineered data and a benchmark analysis of the Berkeley growth data. We conclude our investigation with an application to  time course gene expression.


\section{Introduction}\label{introduction}
Functional data is often characterized by both shape and phase variability. A typical example where
these two sources of variation are clearly identified and interpreted  are data arising from the study of human growth.  Panel (a) and (b) of Figure \ref{growthVelocity} shows growth velocity curves of 39 boys and 54 girls from the Berkeley Growth Study \citep{Tudden:Sny:1954}. An overall pattern is observed that growth velocity decelerates to zero from infancy to adulthood, with some subtle acceleration-deceleration pulses during late childhood and a prominent pubertal growth spurt. In this setting, phase variability is identified as variation in the timing of subject-specific growth. Explicit consideration of phase variability is necessary in order to obtain consistent estimation of typical growth patterns.

The formal analytical treatment of this problem has a long history in Statistics and Engineering. Initial contributions focused on curve alignment (registration) via dynamic time warping \citep{Sakoe:Chiba:1978, Wang:Gass:alig:1997, Wang:Gass:sync:1999} or landmark registration \citep{Gass:Knei:sear:1995}. Model-based alternatives represent subject-specific profiles as a parametric transformation of a common smooth regression function, evaluated over random functionals of time  \citep{Lawt:Sylv:Magg:self:1972, Knei:Gass:conv:1988}. Several of these methods involve a transformation of both the $x$ and $y$ axes, essentially defining the mean profile for curve $i$ as $f_i(x)=b_i + a_i m\big(\mu_i(x)\big) $, where $\mu_i(x)$ is a monotone transformation function accounting for phase variability. In longitudinal settings, \citet{Brum:Lind:self:2004} introduced a mixed effect formulation of these models, formally accounting for dependence within subject. Similary,  \citet{Telesca:Inoue:2007} proposed a Bayesian hierarchical curve registration (BHCR) model allowing for posterior inference on both the shape function $m(\cdot)$ and transformation functions $\mu_i(x)$. Whereas these considerations are valid for any function argument $x$, it is most natural to  think of $x$ as a time scale. In the following, we will therefore focus on the case of functional data observed over time.

Besides technical differences, these models of curve registration share a fundamental assumption, implying that all observed functional profiles are generated through semi-parametric transformations of a common shape  $m(\cdot)$. While this assumption is likely to be warranted in standard applications, the increasing popularity of these methods for the analysis of more general data classes \citep{TelescaEtal2009, TelescaErosheva:2012} motivates a methodological extension, conceiving the possible existence of shape-invariant subgroups, with group shapes $m_1(\cdot),\ldots,m_k(\cdot)$.

We are not the first to recognize a need for combing clustering and registration. Stepwise procedures, where first curves are registered and then clustered according to a chosen heuristic, have been explored in several applications \citep{LiuMuller2004, TangMuller2009, SlaetsEtal2012}.  Joint clustering and registration procedures have been discussed by \citet{GaffneySmyth2005} and \citet{LiuYang2009}. While stepwise procedures are likely to provide suboptimal estimation, available joint clustering and registration techniques have only been developed under the assumption of  linear  shape invariant time transformations, where $\mu_i(t) = \alpha_it + \beta_i$. Furthermore, model complexity, conceived as the number of clusters, is only treated as a nuisance parameter and fixed in a post-hock fashion via BIC or pBIC \citep{ChouReichl1999}.

\vskip.1in
\noindent We extend the BHCR model of \citet{Telesca:Inoue:2007} to allow for shape-subgroups. Our proposal is based on a reproducing kernel  (B-spline) representation of both shape and time transformation functions. To relax the homomorphic assumption, we define a non-parametric prior over shape functionals via a Dirichlet process (DP) mixture \citep{Ferguson1973, Antoniak1974, Quintana2006}. Clustering is achieved implicitly  and is interpreted in  terms of shape similarities. The number of clusters is subject to direct estimation and inferences account fully for this layer of uncertainty, without the need for post-hock adjustments.  Furthermore, we show how posterior simulation remains straightforward via a simple extension to standard Metropolis within Gibbs MCMC transitions.

Following \citet{LiuMuller2004}, we show how this modeling approach is particularly useful for the analysis of time-course gene expression.  While it is known that co-expressed genes are likely to be co-regulated, various regulation mechanisms, such as feedback loops and regulation cascades, may warp the timing of expression for genes involved in the same process or regulatory pathway \citep{Weber2007}. It is therefore desirable to have a model that can assign genes with similar, yet time-warped, expression profiles to the same cluster \citep{QianEtal2001, Qin2006}. In other words, it is important to have a model that is phase-variation tolerant when defining curve subgroups.

The remainder of this article is organized as follows. In section \ref{modelFormulation}, we describe the the sampling model and priors. A posterior simulation strategy via MCMC is described in Section \ref{Posterior}. In section \ref{simulation}, we apply the joint model to simulated datasets and compare it with single-purpose models: a clustering only model and a registration only model. In section \ref{growthData} we apply the model to the Berkeley Growth Study data. In section \ref{fibroResponse}, we apply the model to time course microarray data of response of human fibroblasts to serum stimulation. Finally in section \ref{discussion}, we conclude the paper with a critical discussion.

\section{Model Formulation}\label{modelFormulation}

\subsection{Sampling Model}\label{modelDescription}
Let $y_i(t)$ denote the observation of curve $i$ at time $t$, where $i=1, \ldots, N$ and $t \in T = [t_1, t_n]$.
The sampling model is specified as follows:
\begin{equation}\label{modelStage1}
 y_i(t)=c_i+a_i m_i\{\mu_i(t, \bmath{\phi}_i), \bmath{\theta}_i\} + \epsilon_{i}(t),
\end{equation}
where $\epsilon_{i}(t)\sim \mathrm{N}(0, 1/\tau_i)$ and $\tau_i$ is the precision parameter.

In formula (\ref{modelStage1}), $\mu_i()$ is the curve specific time transformation function,  characterizing the latent time scale of curve $i$, and $m_i()$ is the curve specific shape function. The apparent lack of identifiability between $\mu_i()$ and $m_i()$ will be resolved in \S \ref{Priors} by specifying a random probability functional prior for $m_i()$, implicitly producing functional clusters.

To achieve flexible modeling of both time transformation and shape functions, we use B-splines \citep{deBoor:1978}.
We model
$\mu_i(t, \bmath{\phi}_i)=\bmath{B}^T_{\mu}(t)\bmath{\phi}_i$, where $\bmath{B}_{\mu}(t)$ is the B-spline basis vector at time $t$ and $\bmath{\phi}_i$ is the curve specific basis coefficient vector.
$\mu_i()$ is a monotone function mapping the sampling time interval $T$ to the interval $\mathcal{T}=[t_1-\Delta, t_n+\Delta]$, with expansion constraint $\Delta \geq 0$ to allow the time scale to be transformed outside the observed sampling time interval $T$. To ensure monotonicity and function image boundaries, we impose the following constraints
\begin{equation}\label{phiConstraint}
(t_1-\Delta)\leq \phi_{i1} < \ldots < \phi_{iq} < \phi_{i(q+1)} < \ldots < \phi_{iQ} \leq (t_n+\Delta).
\end{equation}
We model shape functions as
$m_i\{\mu_i(t, \bmath{\phi}_i), \bmath{\theta}_i\}=\bmath{B}^T_m\{\mu_i(t, \bmath{\phi}_i)\} \bmath{\theta}_i$ where $\bmath{B}_m(\cdot)$ is a B-spline basis vector and $\bmath{\theta}_i$ is the curve specific basis coefficient vector. No constraints are usually imposed on $\bmath{\theta}_i$, unless specific shapes are preferred a-priori (see for example, \cite{TelescaErosheva:2012}).

We note that the stochastic functionals $m_i()$ and $\mu_i()$ are now fully described by the distributions of $\bmath{\theta}_i$ and $\bmath{\phi}_i$ respectively. Identifiability of $\bmath{\phi}_i$ is ensured by modeling $\bmath{\theta}_i$ as a Dirichlet process mixture. In this setting, realizations of $\bmath{\theta}_i$ are discrete with probability one, with $K<N$ unique component vectors $\bmath{\theta}_k^*$, ($k=1,\ldots,K$). These component vectors, in turn, define cluster specific shape functions $m_k^*()$, to which member curves are aligned through $\mu_i(t)^{-1}$. Details are discussed in the following section.

\subsection{Prior Model}\label{Priors}
We assume that shape function parameters $\bmath{\theta}_i$ and precisions $\tau_i$
to follow a Dirichlet process mixture prior. Let $G_0()$ be a base distribution absolutely continuous with respect to the Lebesque measure on $\mathbb{R}^p\cup \mathbb{R}^+$  and $\delta(\bmath{\theta}_j, \tau_j)$ a Dirac mass at $(\bmath{\theta}_j, \tau_j)$. Using a predictive P\`{o}lya urn scheme \citep{BlackwellMacQueen1973}, we specify the prior distribution as follows:
\begin{equation}\label{DPprior}
\bmath{\theta}_i, \tau_i|\bmath{\theta}_{-i}, \tau_{-i} \sim \frac{\alpha}{(\alpha+N-1)} G_0(\bmath{\theta}_i, \tau_i)+\frac{1}{(\alpha+N-1)}\sum_{j\neq i}\delta(\bmath{\theta}_j, \tau_j),
\end{equation}
where $-i=\{j: j\neq i\}$ is the set all the indices other than $i$ and $\alpha$
is the weight parameter
of the Dirichlet process model.
This prior generates the shape $\bmath{\theta}_i$ and error precision $\tau_i$ for curve $i$, from a mixture involving a random draw from
the base density $G_0()$ or the point mass $(\bmath{\theta}_j, \tau_j)$'s, $j \neq i$.

Realizations from the prior in (\ref{DPprior}) define a discrete distribution, implying ties among $(\bmath{\theta}_i, \tau_i)$'s, $i=1,\ldots,N$.  These ties are naturally interpreted as clusters among the $N$ curves, namely, curve $i$ and $j$ belong to the same cluster if $(\bmath{\theta}_i, \tau_i)=(\bmath{\theta}_j, \tau_j)$. As a result, only $K < N$ unique values are observed, each of which is associated with a cluster and is denoted by $(\bmath{\theta}_k^*, \tau_k^*)$, $k=1, \ldots, K$. In this setting, we can re-express formula (\ref{DPprior}) as:
\begin{equation}\label{DPprior2}
\bmath{\theta}_i, \tau_i|\bmath{\theta}_k^*, \tau_k^{*} \sim
\frac{\alpha}{(\alpha+N-1)} G_0(\bmath{\theta}_i, \tau_i)+\frac{1}{(\alpha+N-1)}\sum_{k=1}^{K_{-i}}n_{k(-i)}\delta(\bmath{\theta}_k^*, \tau_k^{*}),
\end{equation}
where $n_{k(-i)}$ is the size of cluster $k$ and $K_{-i}$ is number of clusters when curve $i$ is excluded. The representation above implies that a complete sample of $(\bmath{\theta}_i, \tau_i)$, $(i=1,\ldots,N)$ is in one to one correspondence with a set of unique values, $(\bmath{\theta}_k^*, \tau_k^*)$, $(k=1,\ldots,K)$, through cluster labels $\bmath{s}=(s_1, \ldots, s_N)$. Specifically, $s_i=k$ if $(\bmath{\theta}_i, \tau_i)=(\bmath{\theta}_k^*, \tau_k^{*})$ and $s_i=K_{-i}+1$ if $(\bmath{\theta}_i, \tau_i)$ is a new sample from $G_0(\bmath{\theta}_i, \tau_i)$, which means curve $i$ forms a new cluster of its own. As a result, the number of clusters $K$ is also determined implicitly.

We note that if we omit the time transformation modeling with $\mu_i(t, \bmath{\phi}_i)$ and use time $t$ directly in the shape functions $m_i(t, \bmath{\theta}_i)$, our model reduces to  standard functional clustering via Dirichlet process mixtures.

\vskip.2in
\noindent We assume that the base DP mixture density factors as  $G_0(\bmath{\theta}_i, \tau_i)=p(\bmath{\theta}_i \mid \tau_i)p(\tau_i)$,
where $\bmath{\theta}_i \mid \tau_i \sim \mathrm{N}\left(\bmath{0}, (\tau_{\theta}\tau_i\bmath{\Sigma})^{-1}\right)$ and $\tau_i \sim \mathrm{Ga}(a, b)$, a Gamma distribution with mean $a/b$.
The specific form of the precision matrix $\Sigma$ is determined by  a second-order shrinkage process: $\theta_{ip}-\theta_{i(p-1)}=$ $\theta_{i(p-1)}-\theta_{i(p-2)}+\xi_{ip}$ with $\xi_{ip} \sim \mathrm{N}\big(0, 1/(\tau_{\theta}\tau_i)\big)$  ($p=1,\ldots,P$) where $P$ is the dimension of $\bmath{\theta}_i$ and $\theta_{i0}=\theta_{i(-1)}=0$  \citep{Lang:Brez:baye:2004}. In this setting,
the product $\tau_{\theta}\tau_i$ can be interpreted as a smoothing parameter for curve $i$.

Similarly, we also use a penalized B-spline
prior on the time transformation function parameters $\bmath{\phi}_i$. In particular, letting $\bmath{\phi}_0$ be the vector associated with identity transformation so that $\mu(t, \bmath{\phi}_0)=t$, we assume
$\phi_{iq}-\phi_{0q}=$ $\phi_{i(q-1)}-\phi_{0(q-1)}+\nu_{iq}$ with $\nu_{iq} \sim \mathrm{N}(0, 1/\tau_{\phi})$  ($q=1,\ldots, Q$) where $Q$ is dimension of $\bmath{\phi}_i$ and $\phi_{i0}=0$,  implying  $\bmath{\phi}_i \sim \mathrm{N}\left(\bmath{\phi}_0, (\tau_{\phi}\Omega)^{-1}\right)$. In the foregoing, $\Omega$ is deterministic and $\tau_{\phi}$ is interpreted as a smoothing parameter.

Following \citet{TelescaEtal2009}, when modeling cluster specific common shape functions, we let the number of spline knots equal to the number of sampling time points. 
For the curve specific time transformation functions structural smoothness is imposed by their monotonicity (\ref{phiConstraint}), suggesting parsimony in the choice of the number of knots. In many application contexts, $1$ to $4$ equally spaced interior knots allow for enough flexibility
in the representation of time transformation.

\vskip.2in
\noindent For ease of computation, we complete our model with priors and hyperpriors following principles of conditional conjugacy. Specifically, curve specific mean levels parameters are specified as
$c_i \sim \mathrm{N}(c_0, 1/\tau_c)$ and
curve specific amplitude parameters
$a_i \sim \mathrm{N}(a_0, 1/\tau_a)I(a_i>0)$. The assumption of strictly positive amplitudes is appropriate if synchronous but negatively correlated curves are to be clustered separately. Removing positivity restrictions will imply clustering of synchronous profiles.
We complete our prior specifications assuming  $c_0 \sim \mathrm{N}(0, 1/\tau_{c_0})$, $a_0 \sim \mathrm{N}(1, 1/\tau_{a_0})$, $\tau_a \sim \mathrm{Ga}(a_a, b_a)$ and $\tau_c \sim \mathrm{Ga}(a_c, b_c)$. Smoothing parameters priors are specified as
$\tau_{\theta} \sim \mathrm{Ga}(a_{\theta}, b_{\theta})$ and $\tau_{\phi} \sim \mathrm{Ga}(a_{\phi}, b_{\phi})$. Finally, the weight parameter of the Dirichlet process mixture  is  defined as  $\alpha \sim \mathrm{Ga}(a_{\alpha}, b_{\alpha})$.

\section{Posterior  Inference}\label{Posterior}
\subsection{Posterior Simulation}\label{MCMC}
Markov Chain Monte Carlo simulation from the posterior distribution is conceptually straightforward and obtained as a simple sequence of Metropolis-Hastings within Gibbs transitions.

For ease of notation, we let $\bmath{\eta}_i=(\bmath{\theta}_i^t, \tau_i)^t$ and $\bmath{y}_i=(y_i(t_1), \ldots, y_i(t_n))^t$. 
Without loss of generality, we also assume that curves are of the same length $n$. The proposed Markov transition sequence is implemented by: (i) sampling $(\bmath{\phi}_i,\,\bmath{\eta}_i,\,\bmath{s}_i)$ given all other parameters, (ii) resampling $\bmath{\eta}^*_k=(\bmath{\theta}_k^{t*}, \tau_k^{*})^t$ given cluster indicators $\bmath{s}_i$ and all other parameters and (iii) sampling $\alpha$ and remaining parameters from their full conditional posteriors. We outline details as follows.

\vskip.1in
\noindent (i) {\it Sampling $(\bmath{\phi}_i,\,\bmath{\eta}_i,\,\bmath{s}_i)$.} The full conditional posterior of $\bmath{\eta}_i$ is a Dirichlet process mixture  with updated mixing probabilities and components \citep{Escobar1994, WestEtal1994}:
\begin{equation}\label{eta}
\bmath{\eta}_i\mid \bmath{\eta}_k^*, \bmath{\phi}_i, \bmath{y}_i
\sim \frac{q_{i0}G_i(\bmath{\eta}_i \mid \bmath{\phi}_i, \,\bmath{y}_i) +\sum\limits_{k=1}^{K_{-i}}q_{ik}\delta(\bmath{\eta}_k^*)} {q_{i0}+\sum\limits_{k=1}^{K_{-i}}q_{ik}},
\end{equation}
where 
$q_{i0}=\alpha\int f(\bmath{y}_i\mid \bmath{\phi_i}, \bmath{\eta}_i)\mathrm{d}G_0(\bmath{\eta}_i)$ is $\alpha$ times the
 marginal likelihood of $\bmath{y}_i$, 
 $G_i(\bmath{\eta}_i \mid \bmath{\phi}_i, \,\bmath{y}_i) \propto f(\bmath{y}_i\mid \bmath{\phi_i}, \bmath{\eta}_i)G_0(\bmath{\eta}_i) $ is the full conditional density of $\bmath{\eta}_i$ 
 and $q_{ik}=n_{k(-i)}f(\bmath{y}_i \mid \bmath{\phi}_i, \bmath{\eta}_k^*)$ is the product of cluster size $n_{k(-i)}$ and the likelihood associated with $\bmath{\eta}_i=\bmath{\eta}_k^*$.

To improve mixing rates, we combine the sampling of $\bmath{s}_i$, $\bmath{\eta}_i$ and $\bmath{\phi}_i$. Specifically, we use $K_{-i}$ copies of $\bmath{\phi}_i$, $\bmath{\phi}_i^1, \ldots, \bmath{\phi}_i^{K_{-i}}$, one for each cluster. We update $\bmath{\phi}_i^k$, assuming $\bmath{\eta}_i=\bmath{\eta}_k^*$, with the Metropolis-Hastings algorithm as in \citet{Telesca:Inoue:2007}, so that the appropriate time transformation is found for curve $i$ to be registered with the common shape function of cluster $k$. We  calculate each $q_{ik}$ $(k =1, \ldots, K_{-i})$ in (\ref{eta}) using the corresponding $\bmath{\eta}_k^*$ and $\bmath{\phi}_i^k$.  When calculating $q_{i0}$, we use the value of $\bmath{\phi}_i$ from the previous iteration of the Gibbs sampler.

Specifically, let $\bmath{B}_i=\bmath{B}_m\{u_i(\bmath{t}, \bmath{\phi}_i)\} = \bmath{B}_m\{\bmath{B}_{\mu}^T(\bmath{t})\bmath{\phi}_i\}$ and  $\bmath{c}_i=c_i\bmath{1}$, and define the following summaries: $\bmath{E}_i=a_i^2\bmath{B}_i^T\bmath{B}_i+\tau_{\theta}\bmath{\Sigma}$, $\bmath{\mu}_i=a_i\bmath{B}_i^T(\bmath{y}_i-\bmath{c}_i)$, $a'_i=\frac{n}{2}+a$ and $b'_i=\frac{1}{2}(\bmath{y}_i-\bmath{c}_i)^T (\bmath{I}-a_i^2\bmath{B}_i\bmath{E}_i^{-1}\bmath{B}_i^T)(\bmath{y}_i-\bmath{c}_i) +b$. To sample $\bmath{\eta}_i$ from its full conditional (\ref{eta}), we follow the procedure below:
\begin{enumerate}
  \item Sample cluster membership $s_i$ which takes values on $K_{-i}+1, 1, \ldots, K_{-i}$ with probabilities proportional to $q_{i0}, q_{i1}, \ldots, q_{iK_{-i}}$.
  \item If $s_i=K_{-i}+1$, we keep $\bmath{\phi}_i$ unchanged.  Curve $i$ forms a new cluster, and a draw of $\bmath{\eta}_i$ from $G_i(\bmath{\eta}_i \mid \bmath{\phi}_i,\bmath{y}_i)$ is obtained by first sampling $\tau_i \sim \mathrm{Ga}(a'_i, b'_i)$ and then sampling $\bmath{\theta}_i \mid \tau_i \sim \mathrm{N}(\bmath{E}_i^{-1}\bmath{\mu}_i ,(\tau_i\bmath{E}_i)^{-1})$. If $s_i=k$, we  use the corresponding $\bmath{\phi}_i^k$ as a draw of $\bmath{\phi}_i$ and  let $\bmath{\eta}_i=\bmath{\eta}_k^*$.
\end{enumerate}

\vskip.2in
\noindent(ii) {\it Resampling $\bmath{\eta}^*_k$ given cluster indicators $\bmath{s}_i$.} After a sample of $\bmath{\eta}^T=(\bmath{\eta}_1^T, \ldots, \bmath{\eta}_N^T)$ and $\bmath{s}=(s_1, \ldots, s_N)^T$ is generated, to improve mixing rates, we update each $\bmath{\eta}_k^*$  from its full conditional $G_k(\bmath{\eta}_k^*\mid \bmath{y}_{i\in S_k}) \propto \prod_{i\in S_k}\!f(\bmath{y}_i\mid \bmath{\phi_i}, \bmath{\eta}_i)G_0(\bmath{\eta}_k^*)$, where $S_k=\{i:s_i=k\}$ is the set of curves in cluster $k$ \citep{MacEachernMuller1998}. Furthermore, $G_k(\bmath{\eta}_k^*\mid \bmath{y}_{i\in S_k}) \propto p(\bmath{\theta}_k^*\mid\tau_k^*,\bmath{y}_{i\in S_k})p(\tau_k^*\mid\bmath{y}_{i\in S_k})$, s.t.
\begin{equation}\label{etak}
\begin{split}
&\bmath{\eta}_k^*\mid \bmath{y}_{i\in S_k} \sim \mathrm{N}(\bmath{E}_k^{-1}\bmath{\mu}_k, (\tau_k^*\bmath{E}_k)^{-1})\\
&\times \mathrm{Ga}\left(\frac{1}{2}\sum_{i\in S_k}\!\!n_i+a, \frac{1}{2}\left(\sum_{i\in S_k}(\bmath{y}_i-\bmath{c}_i)^T(\bmath{y}_i-\bmath{c}_i) - \bmath{\mu}_k^T\bmath{E}_k^{-1}\bmath{\mu}_k\right)+b\right),
\end{split}
\end{equation}
where $\bmath{E}_k=\sum_{i\in S_k} a_i^2\bmath{B}_i^T\bmath{B}_i + \tau_{\theta}\bmath{\Sigma}$ and $\bmath{\mu}_k=\sum_{i\in S_k}a_i\bmath{B}_i^T(\bmath{y}_i-\bmath{c}_i)$.

\vskip.2in
\noindent(iii) {\it Sampling $\alpha$ and all hyper parameters.}
To develop the full conditional of $\alpha$, we note that $p(K|\alpha,N)\propto N!\alpha^K\frac{\Gamma(\alpha)}{\Gamma(\alpha+N)}$ \citep{Antoniak1974}. Following \citep{West1992}, we define an auxiliary random quantity  $x\mid \alpha \sim \mathrm{B}(\alpha+1, N)$ and
a mixing probability $\pi_x$:
$$\frac{\pi_x}{1-\pi_x}=\frac{a_{\alpha}+K-1}{N(b_{\alpha}-log(x))}.$$
Conditioning on $x$, it is easily shown that the full conditional distribution of $\alpha$ is a mixture of gamma densities.
Specifically,
\begin{equation}\label{alpha3}
\begin{split}
&\alpha\mid x, K \sim\\
&\pi_x \mathrm{Ga}(a_{\alpha}+K, b_{\alpha}-log(x)) + (1-\pi_x)\mathrm{Ga}(a_{\alpha}+K-1, b_{\alpha}-log(x))
\end{split}
\end{equation}
The rest of the model parameters are simulated directly from their full conditional posterior distributions. Detailed results are reported in Web Appendix A.

\subsection{Posterior Inference}\label{inference}
We base our inference on MCMC samples from the posterior distribution of the model parameters. Inference for functional quantities is obtained by post-processing these finite-dimensional posterior samples.
To get a point estimate of the clustering structure we use 
the maximum a-posterior (MAP) clustering.

Given $M$ posterior samples of $\bmath{\phi}_i^{(j)}$, $(j=1, \ldots, M)$, posterior samples of the time transformation function $\mu_i(t)$ at any time point $t \in T$ can be calculated as:
\begin{equation}\label{muPosterior}
\mu_i^{(j)}(t)=\mu_i^{(j)}(t, \bmath{\phi}_i^{(j)})=\bmath{B}_{\mu}^T(t)\bmath{\phi}_i^{(j)}.
\end{equation}
Here, the posterior mean function $\hat{\mu}_i(t)=\frac{1}{M}\sum_{j=1}^M\mu_i^{(j)}(t)$ provides an point estimate of $\mu_i(t)$, and  curves are registered on the transformed time scales $\hat{\mu}_i(t)$ within each cluster.

Similar estimators are defined for cluster-specific shape functions:
\begin{equation}\label{clusterShapeFunction}
m_k(t)=c_0+a_0\bmath{B}^T_m(t)\bmath{\theta}_k^*, \;\;\;\; (k=1,\ldots,K);
\end{equation}
and curve specific profiles:
\begin{equation}\label{curveShapeFunction}
m_i(\mu_i(t))=c_i+a_i\bmath{B}^T_m(\bmath{B}^T_{\mu}(t)\bmath{\phi}_i)\bmath{\theta}_i.
\end{equation}
Point-wise credible intervals and functional bands are easily obtained as empirical quantiles.
Alternatively, the simultaneous credible band for a function $f(\cdot)$ can be obtained as described in \citet{CrainiceanuEtal2007} and \citet{Telesca:Inoue:2007}. 

To assess the model fit, we use the Conditional Predictive Ordinate (CPO) \citep{geisser1979predictive, Pettie1990}. The theoretical CPO for curve $i$ is defined as
\begin{equation}\label{CPO}
    p(\bmath{y}_i|\bmath{y}_{-i}) = \int p(\bmath{y}_i|\Theta)p(\Theta|\bmath{y}_{-i})d\Theta,
\end{equation}
where $\Theta$ denotes the collection of all 
model parameters. A Monte Carlo estimate, based on posterior draws is defined as:
\begin{equation}\label{CPOMCMC}
    \mathrm{CPO}_i= \left\{M^{-1}\sum_{j=1}^M p(\bmath{y}_i|\Theta^{(j)})^{-1}\right\}^{-1}.
\end{equation}
Overall model fit is assessed using the  log pseudo marginal likelihood (LPML), computed as:
\begin{equation}\label{LPML}
    \mathrm{LPML}=\sum_{i=1}^N \log(\mathrm{CPO}_i).
\end{equation}

\section{A Monte Carlo Study of Engineered Data}\label{simulation}
We carry out a simulation study  aimed at assessing the merits of joint clustering and registration and comparing the performance of
our modeling strategy to common clustering techniques.  We consider 100 datasets,
each consisting of,
$45$ curves in $4$ clusters. Each curve $i$ is generated as: $y_i(t)=c_i+a_if_k(\mu_i(t))+\epsilon_{it}$, (if $s_i=k$), with $c_i \sim \mathrm{N}(0, \sigma=0.3)$, $a_i \sim \mathrm{N}(1, \sigma=0.3)I(a_i>0)$ and $\epsilon_{it} \sim \mathrm{N}(0, \sigma=0.3)$. We simulate realizations at $21$ equidistant time points within interval $T=[0,20]$.
We use the following cluster specific
shape functions: $f_1(t)=\cos(t/4)+\sin(t/4)$, $f_2(t)=\cos(t/8)$, $f_3(t)=\sin(t/2)$ and $f_4(t)=0$. Cluster $4$ serves as a noise cluster with no signal.  Finally, time transformations
$\mu_i(t)$ are generated as  a monotone linear combination of B-spline basis, defined by one interior knot at $t=10$.

We fit our model overparametrizing functional forms and fix $31$ equidistant interior knots between $-5$ to $25$ to model common shapes spline bases, and $3$ interior knots at $(5, 10, 15)$ to model time transformation spline bases. Precisions and the Dirichelet mixture weight $\alpha$ are assigned diffuse $\mathrm{Ga}(0.01,0.01)$ priors, (mean=1, variance=100).

\vskip .1in
\noindent To assess the joint model's ability to simultaneous cluster and register curves, we compare the model with a registration only (BHCR) model and the clustering only model as described in section \ref{Priors}. For both the clustering only model and the registration only model, $20K$ iterations is run for the MCMC with the first $10K$ as burn-in.
We also compare our model with model-based clustering (MCLUST) \citep{FraleyRaftery2002a} and functional clustering (FCM) \citep{JamesSugar2003}.

\begin{figure}
\centerline{\includegraphics[width=130mm]{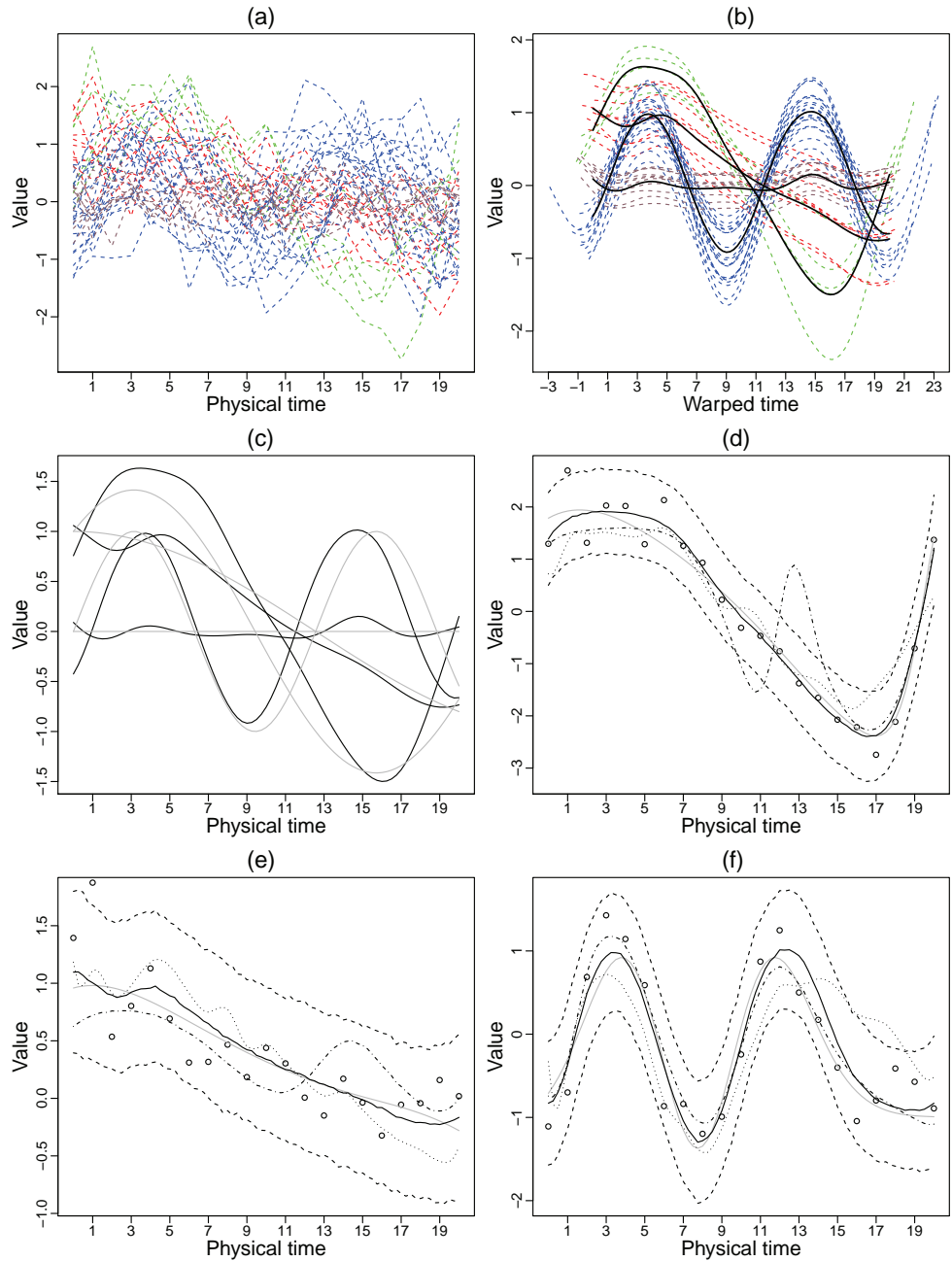}}
\caption{\small \label{simulCurv}\textbf{Simulation study: assessing model fit}.  (a) Simulated unregistered curves in $4$ clusters shown in different colors from one sample dataset. (b) Estimated individuals curves clustered and registered within each cluster, superimposed by the posterior cluster specific common shape functions (solid black). (c) Estimated cluster specific common shape functions (black) and simulation truth (gray). (c)-(f) Three individual curves from cluster 1, 2 and 3, circles indicate the data points for each curve. Estimated individual curves and the true curves are shown in solid black and gray, respectively. $95\%$ simultaneous credible bands are shown as dashed lines. Estimated individual curves from the registration only model and the clustering only model are shown as the dotdash and dotted lines respectively.}
\end{figure}

Figure \ref{simulCurv} shows the results from one of the simulations.  Panel (a) shows $45$ curves color-coded by cluster membership. Panel (b) shows estimated individuals curves clustered and registered within each cluster, with superimposed
cluster-specific shape functions (solid black).
Panel (c) shows posterior expected  cluster-specific shape functions (black) against the simulation truth (gray).   The model is able to accurately recover cluster specific shapes. 
Panel (d)-(f) show results for three individual curves, each from one of the first three clusters. Posterior estimates of individual curves (solid black) are close to the simulation truth (solid gray) and $95\%$ simultaneous credible bands achieve calibrated coverage. 
Also shown are profile estimates from the the registration only model (dotdash) and the clustering only model (dotted). As expected,  since the  registration only model assumes all the curves share a common shape function, cluster-specific functional features are confounded and model fits tend to exhibit spurious features. The clustering only model is inherently highly flexible, as small sub-clusters are allowed to form and fit specific profiles. However, since there tend to be only few curves in each cluster, the loss of information results in noisier estimates.

\begin{figure}
\centerline{\includegraphics[width=130mm]{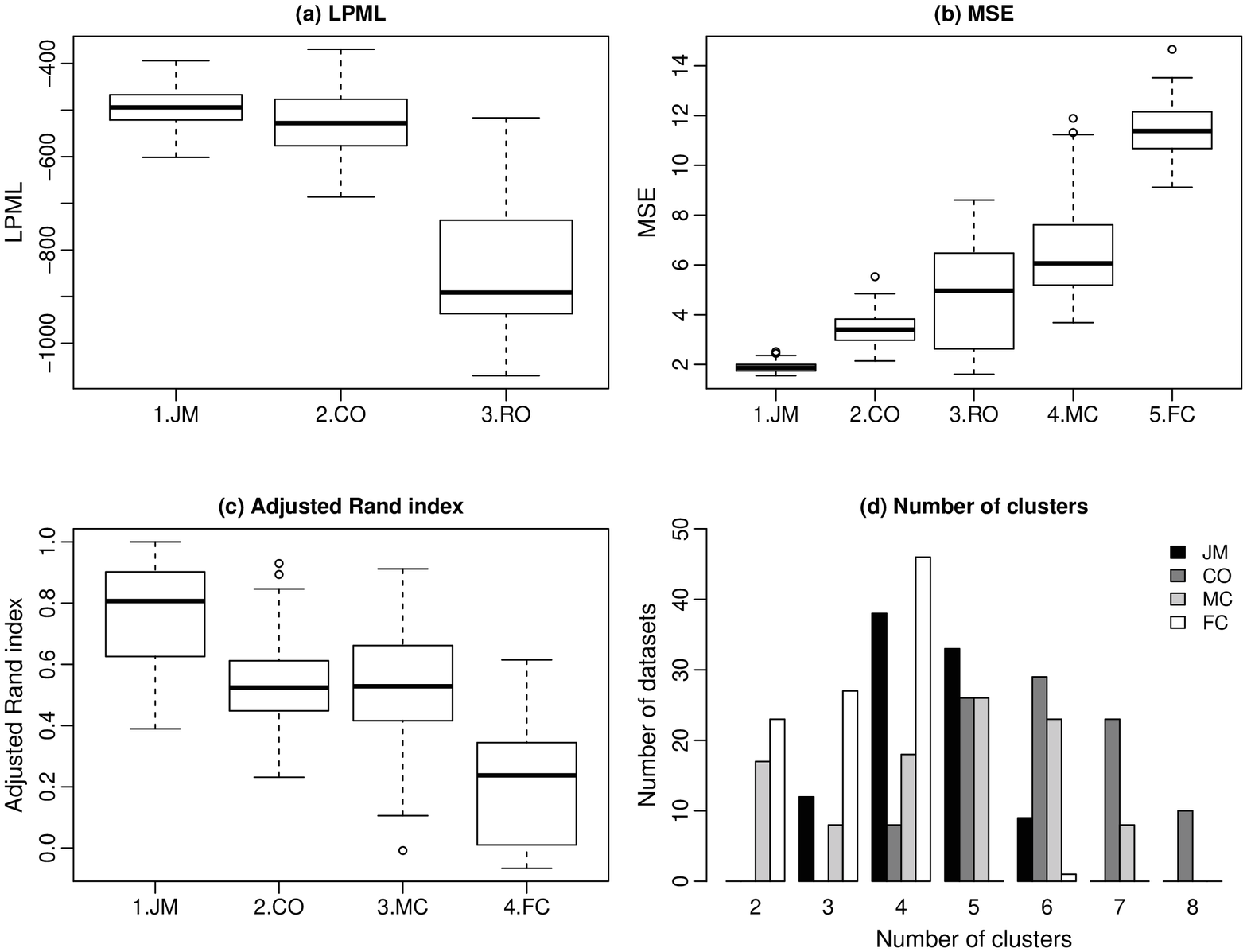}}
\caption{\small \label{simulClust}\textbf{Simulation study: clustering comparison}. (a) Boxplots of log pseudo marginal likelihood (LPML) over the 100 datasets for the joint model (JM), the clustering only model (CO) and the registration only model (RO). (b) Boxplots of MSE over all the estimated curves by the five models, JM, CO, RO, MCLUST (MC) and FCM (FC). (c) Boxplots of adjusted Rand indices for the four clustering models, JM, CO, MC and FC. (d) Bar plots of cluster numbers identified by the four clustering models, JM (black), CO (dark gray), MC (light gray) and FC (white), in the 100 datasets.  }
\end{figure}

\vskip .1in
\noindent We also compared our joint model with MCLUST and FCM in terms of both curve estimation accuracy and clustering accuracy. We apply MCLUST and FCM on the same 100 datasets,
allowing for up to 10 clusters. Figure \ref{simulClust} summarizes comparison results. Panel (a) shows boxplots of the log pseudo marginal likelihood (LPML) comparing the three Bayesian models.
In panel (b) we show boxplots of the simulation mean squared error (MSE) between the estimated and true individual curves 
for all five models. The joint model exhibits best performance in terms of MSE, and the three Bayesian model perform better than MCLUST and FCM. To compare the clustering performance 
we used adjusted Rand index \citep{HubertArabie1985}. Panel (c) shows the boxplots of adjusted Rand indices for the four clustering models. The joint model leads to much higher indices, when compared to the other models considered. The clustering only model does as well as MCLUST and considerably better than FCM. Panel (d) shows a bar plot for the number of clusters identified by the four models. 
Out of the 100 datasets, the joint model identifies 4 clusters in 38 datasets and 5 clusters in 33 datasets. FCM also does well in identifying the correct cluster numbers, specifically it identifies 4 clusters in 46 datasets and 3 clusters in 27 datasets. The clustering only model and MCLUST tend to overestimate the number of clusters.

\vskip .1in
\noindent  We repeated the joint clustering and registration analysis under several prior specifications, in order to assess sensitivity. While the formal task is daunting, due to the large number of parameters in the model, we have found that reasonable variations in prior choice has little impact on final inference, detailed results are reported in Web Appendix A. Clearly, different considerations may apply under different sample size scenarios.

\section{A Cluster Analysis of the Berkeley Growth Data}\label{growthData}
\begin{figure}
\centerline{\includegraphics[width=130mm]{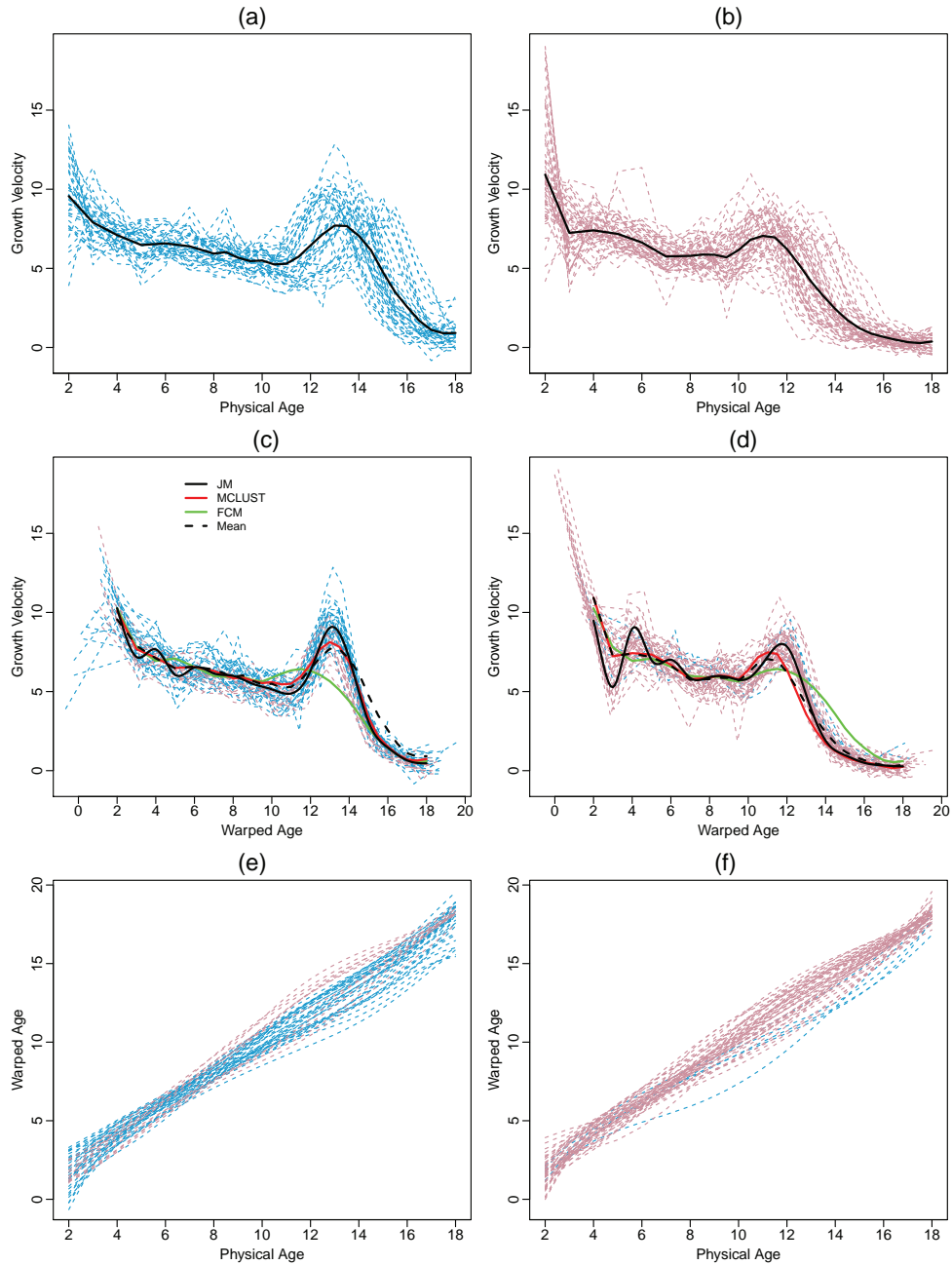}}
\caption{\small \label{growthVelocity}\textbf{Growth velocity data analysis}.
(a) and (b) Individual unregistered growth velocity curves for  39 boys (blue dashed) and 54  girls (pink dashed): cross-sectional means in solid-black. (c) Registered curves in the 1st cluster: 34 boys (blue dashed) and 11 girls (pick dashed). Estimated common shapes are indicated in (solid-black), MCLUST (red), FCM (green) and the cross sectional mean in (dashed-black). (d) Registered curves in the 2nd cluster:  43 girls (pink dashed) and 5 boys (blue dashed). Common shape functions as in (c). (e) and (f) Estimated curve specific time transformation functions for the two clusters: boys (dashed-blue)  and girls (dashed-pink). }
\end{figure}

We apply the proposed model to the well known Berkeley Growth data and compare it with the clustering only model, registration only model, MCLUST and FCM. As discussed in Section \ref{introduction}, the Berkeley Growth Study \citep{Tudden:Sny:1954} recorded the height of 39 boys and 54 girls for 27 time points between age 2 to 18, with one measurement a year before age 9 and two measurements a year after. To construct the growth velocity curves from the original growth curves, a smoothing spline model was fitted to each growth curve, and the first degree derivatives were obtained from the model and used in our comparisons. In Figure \ref{growthVelocity}(a) and (b), the growth velocity curves of boys (blue) and girls (pink) are plotted against age with superimposed cross-section means (black). Within each sex, curves have similarities in shape, while each curve shows individual time and amplitude variation. As pointed out by \citet{Rams:Li:curv:1998} and \citet{Gerv:Gass:self:2004}, failing to account for time variability, produces inconsistent estimates of sex-specific growth velocities. Our analysis is non-standard, as we use sex as a hidden label to assess clustering performance. While illustrative, this exercise finds justification in the fact that sex is expected to explain a large portion of variation in adolescent growth patterns.

Shape functions basis are constructed fixing $\Delta=7$ and placing 27 equidistant interior knots between $-3$ to $23$. To model time transformation functions, we place four interior knots at $(5.2, 8.2, 11.6, 14.8)$ and partition the interval $T=[2, 18]$ into five subintervals. Priors on precisions and mixture weight are set as in Sec. \ref{simulation}. Our inferences are based on $20K$ MCMC iterations, with $10K$ burn-in.

\noindent The model identifies two clusters, seemingly discriminative according to sex.
If we label the first cluster as the "boy" cluster and the second cluster as the "girl" cluster, then $43$ out of $54$ girls are clustered correctly and $34$ out of $39$ boys are clustered correctly. The overall classification accuracy is $83\%$. Estimated time transformation functions, common shape functions and registered curves are shown in Figure \ref{growthVelocity}. Panel (c) and (d) show the registered curves for the $2$ clusters, superimposed by their corresponding common shape functions from the joint model (black solid), MCLUST (red), FCM (green) and the cross sectional mean curves (black dashed). Individual curves are colored by their true gender information, blue for boys and pink for girls. Therefore, pink curves in panel (c) and blue curves in panel (d) show the misclassified cases. Panel (e) and (f) show the estimated curve specific time transformation functions for the two clusters.

\begin{figure}
\centerline{\includegraphics[width=130mm]{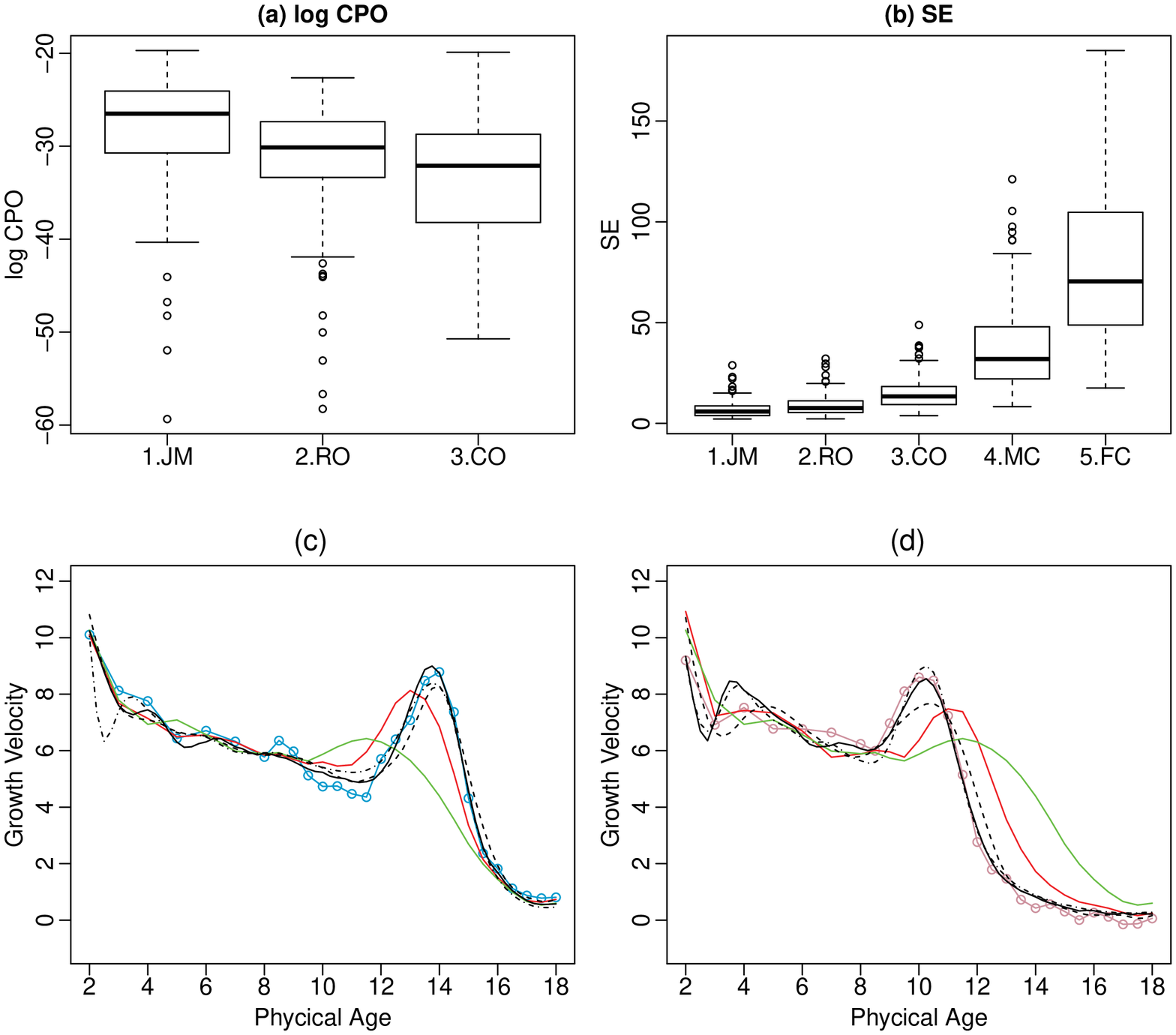}}
\caption{\small \label{growthCompPlot}\textbf{Growth velocity data model comparison}. (a) Boxplots of log CPO of the $93$ individual growth curves by the joint model (JM), the clustering only model (CO) and the registration only model (RO). (b) Boxplots of the squared error (SE) over all the growth curves for JM, CO, RO, MCLUST (MC) and FCM (FC). (c) Model fitting results of the growth velocity curve of a boy (blue line with circles) by JM (black solid), CO (black dashed), RO (black dotdash), MCLUST (red) and FCM (green). (d)Model fitting results of the growth velocity curve of a girl (pick line with circles) by JM (black solid), CO (black dashed), RO (black dotdash), MCLUST (red) and FCM (green). }
\end{figure}

We compared the joint model with the clustering only model, the registration only model, MCLUST and FCM, and the results are shown in Figure \ref{growthCompPlot}. Panel (a) shows the boxplots of CPO of the $93$ individual growth curves by the three Bayesian models. It shows that the joint model fits the data best, followed by the registration only model and the clustering only model. When the curves are not too dramatically different, the registration only model can fit the data accurately by finding a common shape function representing all the curves well. 

As a comparable measure of model fit we 
compute the squared error (SE) between each curve and its fitted profiles.
Panel (b) shows the boxplots of the SE over all the growth curves for the five models. The joint model gives the smallest SE, and the three Bayesian models fit the data better than MCLUST and FCM in terms of SE. Panel (c) and (d) show the model fitting results of two individual curves of a boy and a girl.

\noindent Interpreting sex as a clustering lable, we compare the joint model, the clustering only model, MCLUST and FCM using adjusted Rand indices (RIs). We find the following: FMC (RI = 0.61), MCLUST (RI = 0.47), JM (RI = 0.43) and CO (RI = 0.10). 
By this measure FCM and MCLUST 
seem to outperform our joint  clustering and registrations model, with FCM giving the best clustering results. 
We note that when fitting MCLUST, we set the candidates of cluster numbers to be between 2 to 10, because when 1 is included as a candidate, MCLUST chooses it as the optimal cluster number, which leads to an adjusted Rand index of 0. 
On the other hand, as shown in panel (a) and (b) in Figure \ref{growthVelocity}, FMC and MCLUST seem to provide unsatisfactory estimates
of cluster specific shape functions, which the joint clustering and registration model estimates consistently with the findings of \citet{Rams:Gass:1995} and \citet{Telesca:Inoue:2007}, 
supporting the existence of the mid growth spurts.

\section{Response of Human Fibroblasts to Serum}\label{fibroResponse}
In this section, we apply the joint model to time course expression data of the response of human fibroblasts to serum in a microarray experiment of $8613$ genes \citep{IyerEtal1999}. For human fibroblasts to proliferate in culture, they require growth factors provided by fetal bovine serum (FBS). In their study, after inducing primary cultured human fibroblasts to enter a quiescent state by serum deprivation for $48$ hours, the authors stimulated fibroblasts by adding medium containing $10\%$ FBS. A microarray experiment was then conducted to measure temporal gene expression levels at $12$ time points, from $15$ minutes to $24$ hours after serum stimulation. Furthermore, they selected $517$ genes with substantial time course expression change in response to serum and formed clusters using K-means clustering \citep{EisenEtal1998}. In our analysis, we consider a subset of $78$ genes, since they are associated with clear biological function categories as described in the original paper, and this provides a standard for us to validate the biological relevance of the clustered identified by our model.

We use the same prior setup as in previous sections. To model shape functions we use a maximum expansion constraint $\Delta=6$ and place interior knots at the sampling time points and $5$ equidistant points in two intervals from $-5$ to $-1$ and from $25$ to $29$ respectively. To estimate the time transformation functions, we place four interior knots at $(0.5, 2, 8, 16)$ in the sampling interval $T=[0, 24]$. Our inferences are based on $20K$ MCMC iterations, with $10K$ burn-in.

\begin{figure}
\centerline{\includegraphics[width=130mm]{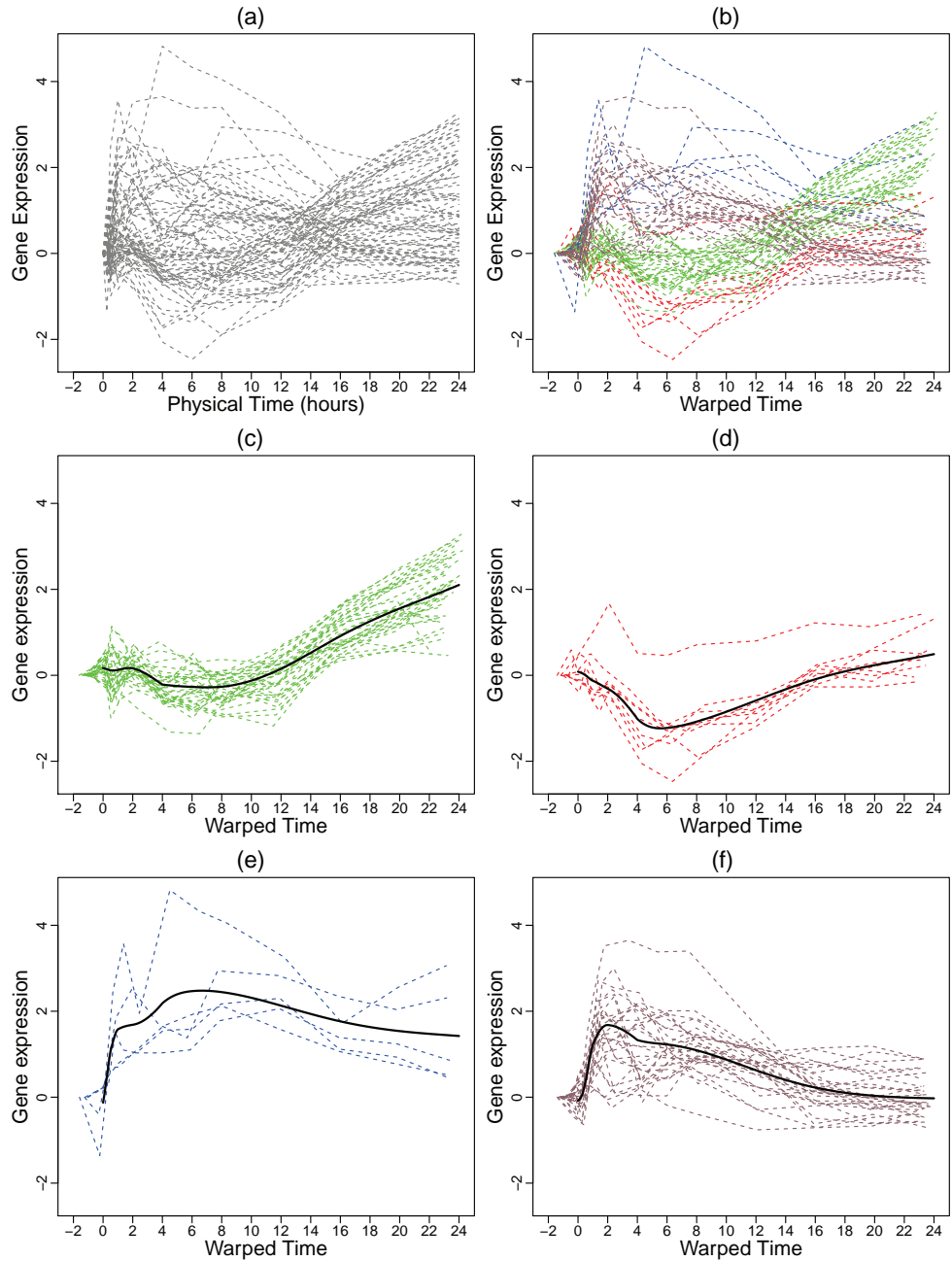}}
\caption{\small \label{geneCurves}\textbf{Human fibroblast gene expression analysis}. (a) Unregistered time coures expression curves for 78 genes selected from a microarray experiment of Human fibroblasts' response to serum. (b) Registered expression curves forming $4$ clusters colored by green, red, blue and pink. (c) Thirty five genes in cluster $1$ superimposed by the cluster specific common shape function (solid black). (d) Nine genes in cluster $2$. (e) Five genes in cluster $3$. (f) Twenty nice genes in cluster $4$.}
\end{figure}

Panel (a) of Figure \ref{geneCurves} shows the unregistered temporal expression curves of the 78 genes selected from the microarray experiment of human fibroblasts' response to serum. Panel (b) shows the registered expression curves which are clustered into $4$ groups. Panel (c)-(f) show the $4$ clusters of registered expression curves separately, superimposed by their cluster specific common shape functions.

\begin{table}
\caption{\label{geneTable} Clusters of genes and their biological functions}
\begin{tabular}{cccc}\hline\hline
Cluster & Size & Typical Genes & Functions  \\ \hline
   1    &  35  & PCNA, Cyclin A, Cyclin B1 & Cell cycle and proliferation \\
          &      & CDC2, CDC28 kinase      &    \\
   2    &  9   & LBR & Cell cycle and proliferation    \\
   3    &  5   & PAI1, PLAUR, ID3 &   Coagulation and hemostasis \\     &    &      & Transcription factors\\
   4    &  29  & MINOR, JUNB, CPBP & Signal transduction         \\
        &      & TIGF, SGK, NET1   & Transcriptional factors     \\ \hline\hline
\end{tabular}
\end{table}

As shown in Figure \ref{geneCurves} (c), genes in cluster $1$ are down-regulated at first and reach their lowest expression levels between $4$ and $12$ hours after serum stimulation. They begin to express about $16$ hours after the serum treatment, which is also the time when the stimulated fibroblasts replicate their DNA and reenter into the cell-division cycle. Several genes in cluster $1$ are known to be involved in mediating cell cycle and proliferation, for instance, PCNA, Cyclin A, Cyclin B1, CDC2 and CDC28 kinase, as shown in Table \ref{geneTable}. Cluster $2$ in Figure \ref{geneCurves}(d) shows similar expression pattern to cluster $1$, except they expression level are lower than those in cluster $1$ through the time window. Genes in cluster $2$ are also involved in cell cycle and proliferation, such as LBR. Figure \ref{geneCurves}(e) shows that genes in cluster $3$ respond immediately to serum stimulation, reach their expression peaks around 10 hours later and remain induced towards the end. They are known to be transcription factors and involved in coagulation and hemostasis because of fibroblasts' role in clot remodeling. Typical genes include PAI1, PLAUR and ID3. As shown in Figure \ref{geneCurves}(f), genes in cluster $4$ are also induced quickly by serum treatment, reach their peaks at about 2 hours, and then gradually return to a quiescent state. Several of the genes here are known to encode transcriptional factors and other proteins involved in signal transduction, such as MINOR, JUNB, CPBP, TIGF, SGK and NET1.

\section{Discussion}\label{discussion}
We propose a Bayesian hierarchical model for joint curve registration and clustering. Compared to previous methods, our proposal comes with several advantages. First, 
the model provides flexible nonlinear modeling for both components of variation.  The Dirichlet process mixture prior over shape functionals strikes an automatic balance between complexity and parsimony. The implied posterior identifies subgroups of homomorphic curves, without the need to specify the number of clusters a priori.  Finally, the increased model flexibility is still  amenable to straightforward posterior simulation via MCMC, which provides exact inferences about a rich set of quantities of interest, without the need for simplifications or approximations.

The proposed B-spline representation of both  shape and time transformation functions requires the a priori specification of the number and placement of spline knots. Our experiences suggests that a set of knots reproducing the original sampling time points works well for shape functions and 1 to 4 equidistant interior knots are enough for time transformation functions, as they carry smoothing properties through  monotonicity constraints. Our simulation study shows that the model is robust to different prior choices. We however maintain, that different considerations may apply to ultra-sparse or, conversely, ultra-dense data settings.

The proposed modeling strategy has potentially broad applications to functional data analysis; especially when curve registration and clustering are of joint interest, as shown in our applications. In the first case study of the Berkeley Growth Data, our model is able to accurately separate growth curves into two clusters labelled by sex, and to correctly estimate the overall growth patterns for both sexes after registering curves in each cluster. In the second case study of time course expression data of human fibroblasts' response to serum, our model identifies fours clusters of genes involved in distinct biological functions.

The proposed estimator of the clustering structure is the MAP clustering. Because Dirichlet process mixtures fully  account for stochasticity in the potential alternative assignment of individual profiles to functional groups, it is possible, in principle, that the clustering structure with the second highest posterior probability is only a little less probable than the MAP clustering, yet it provides quite a different grouping structure.

We have not detected this type of phenomenon  in our analyses. However, when it happens, some care is needed in summarizing complex posterior evidence. A possible alternative strategy to MAP is based on the estimation of a 
pairwise probability matrix whose elements are estimated probabilities that two curves are in the same cluster. Such a matrix can be easily generated by averaging the sampled association matrices from the MCMC output. Elements of an association matrix takes values $1$, if two corresponding curves are in the same cluster, and $0$ otherwise. Hierarchical clustering may be used subsequently as a way to explore grouping structures \citep{MedvedovicSivaganesan2002}. Alternatively, \citet{Dahl2006} proposed a least squares clustering by selecting the sampled clustering which minimizes the sum of squared deviations of its association matrix from the pairwise probability matrix.

Finally, when covariate information is available, the proposed model is easily extended to include a dependent Dirichlet process prior,  using covariates to inform clustering.


\section*{Supplementary Materials}
Supplementary information is available from the authors.

\bibliographystyle{plainnat}
\bibliography{jointModel}

\label{lastpage}

\end{document}